\documentclass[journal,11pt,draftclsnofoot,onecolumn]{IEEEtran}

\usepackage{amsfonts}
\usepackage{amsmath}
\usepackage{amssymb}
\usepackage{amsthm}
\usepackage{cite}
\usepackage{dsfont}
\usepackage{latexsym}
\usepackage{subfigure}
\usepackage{times}
\usepackage{url}
\usepackage{verbatim}

\usepackage{graphicx}

\def\bb0{{\mathbb{0}}}

\def\bb{{\mathbf{b}}}

\def\bm{{\mathbf{m}}}

\def\b0{{\mathbf{0}}}

\def\sf0{{\mathsf{0}}}

\def\rm0{{\mathrm{0}}}

\usepackage[]{hyperref}
\usepackage{hypcap}
\usepackage{enumitem}
\usepackage[usenames, dvipsnames]{color}

\newcommand{\comm}{\mathrm{c}}

\newcommand{\dmse}{\mathrm{DMSE}}

\newcommand{\me}{\mathrm{e}}
\newcommand{\eff}{\mathrm{eff}}

\newcommand{\jm}{\mathrm{j}}

\newcommand{\mcom}{\comm}

\newcommand{\mmse}{\mathrm{MMSE}}

\newcommand{\Ndwel}{N_\mathrm{CPI}}

\newcommand{\mrx}{\mathrm{RX}}

\newcommand{\Tr}[1]{\mathrm{Tr}\left[ #1 \right]}

\newcommand{\txm}{\mathrm{TX}}

\usepackage{graphicx}
\usepackage{setspace,xparse}
\usepackage{xcolor}
\usepackage{bm}

\title{Toward Millimeter Wave Joint Radar-Communications: A Signal Processing Perspective}

\author{
    \IEEEauthorblockN{Kumar Vijay~Mishra\IEEEauthorrefmark{1}\IEEEauthorrefmark{2}\IEEEauthorrefmark{3}, Bhavani Shankar~M. R.\IEEEauthorrefmark{3}, Visa~Koivunen\IEEEauthorrefmark{4}, Bj\"{o}rn~Ottersten\IEEEauthorrefmark{3}\IEEEauthorrefmark{5}, and Sergiy~A.~Vorobyov\IEEEauthorrefmark{4}
    }\\
    \IEEEauthorblockA{\IEEEauthorrefmark{1}IIHR - Hydroscience \& Engineering, The University of Iowa, Iowa City, IA 52246 USA}\\
    \IEEEauthorblockA{\IEEEauthorrefmark{2}Hertzwell, Singapore 059911}\\
    \IEEEauthorblockA{\IEEEauthorrefmark{3}SnT, University of Luxembourg 1855 Luxembourg}\\
    \IEEEauthorblockA{\IEEEauthorrefmark{4}Department of Signal Processing and Acoustics, Aalto University, Espoo 02150 Finland}\\
    \IEEEauthorblockA{\IEEEauthorrefmark{5}KTH Royal Institute of Technology, Stockholm 11428 Sweden}
    \\
    Email: kumarvijay-mishra@uiowa.edu, bhavani.shankar@uni.lu, visa.koivunen@aalto.fi, bjorn.ottersten@uni.lu, sergiy.vorobyov@aalto.fi
}

\begin{document}

\maketitle

\IEEEpeerreviewmaketitle
\begin{abstract}
Synergistic design of communications and radar systems with common spectral and hardware resources is heralding a new era of efficiently utilizing a limited radio-frequency spectrum. Such a joint radar-communications (JRC) model has advantages of low-cost, compact size, less power consumption, spectrum sharing, improved performance, and safety due to enhanced information sharing. 
Today, millimeter-wave (mm-wave) communications have emerged as the preferred technology for short distance wireless links because they provide transmission bandwidth that is several gigahertz wide. This band is also promising for short-range radar applications, which benefit from the high-range resolution arising from large transmit signal bandwidths. Signal processing techniques are critical in implementation of mmWave JRC systems. Major challenges are joint waveform design and performance criteria that would optimally trade-off between communications and radar functionalities. Novel multiple-input-multiple-output (MIMO) signal processing techniques are required because mmWave JRC systems employ large antenna arrays. There are opportunities to exploit recent advances in cognition, compressed sensing, and machine learning to reduce required resources and dynamically allocate them with low overheads. This article provides a signal processing perspective of mmWave JRC systems with an emphasis on waveform design.
\end{abstract}

\section{Introduction}
In recent years, sensing systems (radar, lidar or sonar) that share the spectrum with wireless communications (radio-frequency/RF, optical or acoustical) and still operate without any significant performance losses have captured significant research interest \cite{bliss_RF_convergece_17,Hassan16a}. The interest in such spectrum sharing systems is largely because the spectrum required by the wireless media is a scarce resource, while performance of both communications and remote sensing systems improves by exploiting wider spectrum. In this article, we focus on RF spectrum sharing between radar and communications.

Several portions of frequency bands - from Very High Frequency (VHF) to Terahertz (THF) - are allocated exclusively for different radar applications \cite{cohen2017spectrum}. Although a large fraction of these bands remains  underutilized,  
radars need to maintain constant access to these bands for target sensing and detection as well as obtain more spectrum to accomplish missions such as secondary surveillance, multi-function integrated RF operations, communications-enabled autonomous driving and cognitive capabilities. On the other hand, the wireless industry's demand for spectrum continues to increase for providing new services and accommodating massive number of users with high data rate requirement. The present spectrum is used very inefficiently due to its highly fragmented allocation. Emerging wireless systems such as commercial Long Term-Evolution (LTE) communications technology, fifth-generation (5G), WiFi, Internet-of-Things (IoT), and Citizens Broadband Radio Services (CBRS) already cause spectral interference to legacy military, weather, astronomy, and aircraft surveillance radars \cite{bliss_RF_convergece_17,cohen2017spectrum}. 
Similarly, radar signals in adjacent bands leak to spectrum allocated for communications and deteriorate the service quality. Therefore, it is essential and beneficial for radar and communications to develop strategies to simultaneously and opportunistically operate in the same spectral bands in a mutually beneficial manner.

The spectral overlap of centimeter-wave (cmWave) radars with a number of wireless systems at 3.5 GHz frequency band led to 2012 U. S. President's Council of Advisors on Science and Technology (PCAST) report on spectrum sharing \cite{pcast2012realizing} and changes in regulation for this band became a driver for spectrum sharing research programs of multiple agencies \cite{cohen2017spectrum}.  
Today, 
it is the higher end of the RF spectrum, i.e., the millimeter-wave (mmWave), formally defined with the frequency range 30-300 GHz, that requires concerted efforts for spectrum management because its technologies are in an early development stage. 
Increasingly , the mmWave systems \cite{RapMacSam:Wideband-Millimeter-Wave-Propagation:15} are the preferred technology for near-field communications since it provides transmission bandwidth that is several GHz wide and currently unlicensed. This enables applications which require huge data rates such as 5G wireless backhaul, uncompressed high definition (HD) video, in-room gaming, intra-large-vehicle communications, inter-vehicular communications, indoor positioning systems, and IoT-enabled wearable technologies \cite{daniels200760}. There is also a spurt of novel sensing systems in the mmWave band. Although these devices typically have short ranges because of heavy attenuation by physical barriers, weather, and atmospheric absorption, they provide high range resolution resulting from the wide bandwidth. Typical mmWave radar applications include autonomous vehicles \cite{dokhanchi2019mmwave},  
gesture recognition \cite{lien2016soli}, cloud observation \cite{mishra2018deep}, RF identification \cite{decarli2014novel}, indoor localization \cite{mishra2017sub}, and health monitoring \cite{fortino2012bodycloud}. We now explain the distinct features and JRC challenges of mmWave channel.

\section{The mmWave Channel}
\label{sec:mmw_ch}

Compared to cmWave, the channel environment for mmWave is characterized by unique challenges that motivate the ensuing specific design constraints.
\paragraph{Strong Attenuation} Compared to sub-6 GHz transmissions envisaged in 5G, mmWave signals encounter a more complex propagation environment characterized by higher scattering, severe penetration losses, and lower diffraction. These losses result in mmWave communications links being near line-of-sight (LOS) with fewer non-line-of-sight (NLOS) clusters and smaller coverage areas. Similarly, lower diffraction results in poorer coverage around corners. High attenuation also implies that mmWave radars are useful only at short ranges and, as a result, multipath is a less severe problem.
\paragraph{High Path-Loss and Large Arrays} 
Quite naturally, the mmWave signals suffer from higher path-loss for fixed transmitter (TX) and receiver (RX) gains
. By Friis transmission formula, compensating for these losses while keeping the same effective antenna aperture (or increasing the gain) imposes constraints on the transceiver hardware. 
Since the received power is contingent on the beams of the transmitter and receiver being oriented towards each other, same aperture is accomplished by using steerable antenna arrays whose elements are spaced by at most half the wavelength ($\lambda/2$) of the transmitted signal to prevent undesirable grating lobes
. This inter-element spacing varies between $0.5$-$5$ mm for mmWave carriers. Such narrow spacings impact the choice of RF and intermediate frequency (IF) elements because they should fit in limited space available and precise mounting may be difficult in, for instance, vehicular platforms.
\paragraph{Wide bandwidths} The unlicensed, wide mmWave bandwidth enables higher data rates for communications as well as the range resolution in radar. In automotive radar, this ensures detection of distinct, informative micro-motions of targets such as pedestrians and cyclists \cite{mishra2019doppler}. 
The mmWave receivers sampling at Nyquist rate require expensive, high-rate analog-to-digital converters (ADCs). Large bandwidths also imply that use of low-complexity algorithms in transmitter and receiver processing is critical \cite{dokhanchi2019mmwave}. Further, mmWave channels are sparse in both time and angular dimensions - a property exploited for low-complexity, low-rate reconstruction using techniques such as compressed sensing \cite{mishra2017sub,mishra2018sub}. It is crucial to consider if relevant narrowband assumptions hold in a mmWave application; otherwise, the signal bandwidth is very broad with respect to the center frequency and the steering vectors become frequency-dependent.
\paragraph{Power Consumption} The power consumption of an ADC increases linearly with the sampling frequency. 
At baseband, each full-resolution ADC consumes $15$-$795$ mW at $36$ MHz-$1.8$ GHz bandwidths. In addition, power consumed by other RF elements such as power amplifiers and data interface circuits in conjunction with the narrow spacing between antenna elements renders it infeasible to utilize a separate RF-IF chain for each element. Thus, a feasible multi-antenna TX/RX structure and beamformers should be analog or hybrid (wherein the potential array
gain is exploited without using a dedicated RF chain per antenna and phase shifter) \cite{mendez2016hybrid} 
because fully digital beamforming is infeasible. 
\paragraph{Short Coherence Times} The mmWave environments such as indoor and vehicular communications are highly variable with typical channel coherence times of nanoseconds \cite{RapMacSam:Wideband-Millimeter-Wave-Propagation:15}. 
The reliability and coverage of dynamic mmWave vehicular links are severely affected by the use of narrow beams. The intermittent blockage necessitates frequent beam re-alignment to maintain high data rates. 
Also, mmWave radar requires wide Doppler range to detect both fast vehicles and slow pedestrians \cite{mishra2019doppler}. Short coherence times impact the use of feedback and waveform adaptation in many JRC designs, where the channel knowledge may be invalid or outdated when transmit waveform optimization takes place. 

We now present details of channels models commonly used in mmWave communications and radars.

\subsection{Communications Channel}
Consider a transmitter that employs an antenna array or a single directional antenna with carrier frequency $f$ and TX (RX) antenna gain $G_\txm$ ($G_\mrx$). The LOS communications channel with a delay spread comprising $L_\comm-1$ delay taps is $h_\comm(t,f) = G_\comm \sum_{\ell=0}^{L_\comm-1} \alpha_\ell \me^{- j 2 \pi \tau_\ell f}\me^{j 2 \pi \nu_\ell t}$, where $G_\comm$ is the large-scale communications channel gain at the reception, and $\alpha_\ell$ is the path loss coefficient of the $l^{\text{th}}$ path with time delay $\tau_\ell$ and Doppler shift $\nu_\ell$. The free space attenuation model yields $G_\comm = \frac{G_\txm G_\mrx  \lambda^2 } {(4 \pi)^2 \rho_\mcom^\gamma}$, where $\gamma$ is path loss (PL) exponent 
. Further, $\gamma\approx2$ for mmWave LOS outdoor urban \cite{RapMacSam:Wideband-Millimeter-Wave-Propagation:15} and rural scenarios \cite{MacSunRap:Millimeter-Wave-Wireless:16}. 

\subsection{Radar Channel}
The doubly selective (time- and frequency-selective) mmWave radar channel is modeled after TX/RX beamforming using virtual representation obtained by uniformly sampling in range dimension \cite{KumChoGon:IEEE-802.11ad-based-Radar::17}. 
Assume $L$ uniformly sampled range bins and that the $\ell$-th range bin consists of a few, (say) $K_\ell$, virtual scattering centers. Each $(\ell,k)$-th virtual scattering center is characterized by its distance $\rho_\ell$, delay $\tau_\ell$, velocity $v_{\ell,k}$, Doppler shift $\nu_{\ell,k} = 2 v_{\ell,k}/\lambda$, large-scale channel gain $G_{\ell,k}$, and small-scale fading gain $\beta_{\ell,k}$. Then, the multi-target radar channel model is
$h_r (t,f) = \sum_{\ell=0}^{L-1} \sum_{k=0}^{K_{\ell}-1} G_{\ell,k} \beta_{\ell,k} \me^{- j 2 \pi \tau_\ell f} \cdot \me^{- \jm 2\pi \nu_{\ell,k} t}$. 
The large-scale channel gain corresponding to the $(\ell,k)$-th virtual target scattering center is
$G_{\ell,k} = \frac{\lambda^2 \sigma_{\ell,k}}{64 \pi^3 \rho_\ell^4}$, 
where $\sigma_{\ell,k}$ is corresponding scatterer's radar cross section (RCS). 
The small scale gain is assumed to be a superposition of a complex Gaussian component and a fixed LOS component leading to Rician fading. Similarly, the corresponding frequency selective models can also include Rician fading. They capture, as a special case, the spiky model used in prior works on mmWave communications/radar. In this case, the corresponding radar target models are approximated by the Swerling III/IV scatterers \cite{skolnik2008radar}. 

Further, clustered channel models can be considered to incorporate correlations and extended target scenarios although they remain unexamined in detail. For instance, the conventional mmWave automotive target model assumes a single non-fluctuating (i.e., constant RCS) scatterer based on the Swerling 0 model. 
This greatly simplifies the development and analysis of receive processing algorithms and tracking filters \cite{dokhanchi2019mmwave}. However, when the target is located within the close range of a high-resolution radar, the received signal is composed of multiple reflections from different parts of the same object. This \textit{extended} target model is more appropriate for mmWave applications and may also include correlated RCS \cite{mishra2019doppler}. 

It is typical to assume a frequency-selective Rayleigh fading model for both communications and radar channels during the dwell time comprising $\Ndwel$ coherent processing intervals (CPI). In radar terminology, this corresponds to Swerling I/II target models. In each CPI with $M$ frames, the channel amplitude of each tap is considered to be constant, i.e., a block fading model is assumed. Moreover, constant velocity and quasi-stationarity conditions are imposed on the target model.

\subsection{Channel-Sharing Topologies}
\label{subsec:jrc_top}
 \begin{figure}[t]	
 		\centering
 \includegraphics[width=1.00\textwidth]{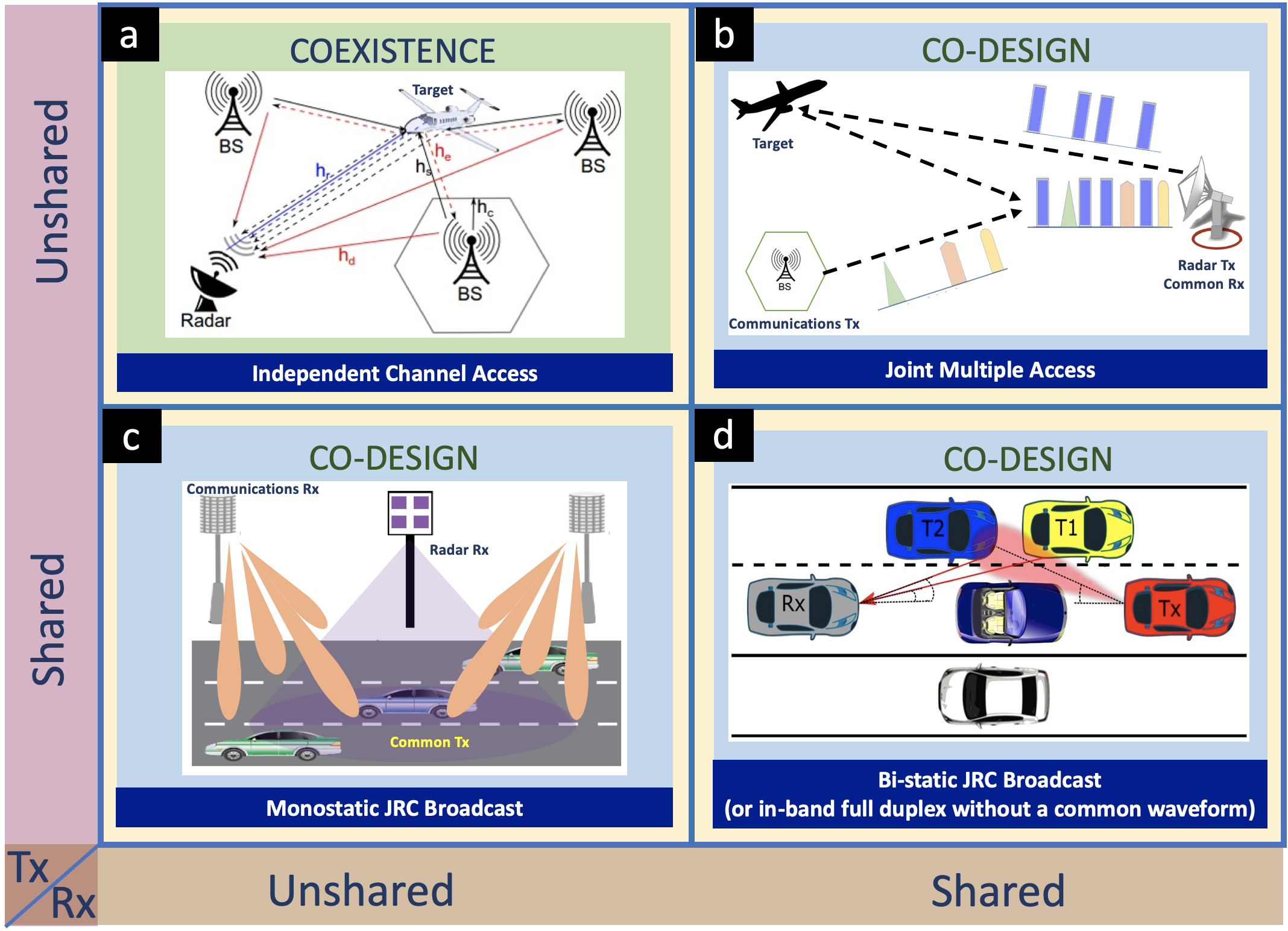}
 		\caption{(a) Spectral coexistence system where radar and communications subsystems are independently located and access the associated radio channels such as radar target channel $h_r$, communications channel $h_c$, radar-to-communications interference $h_s$, and communications-to-radar interference $h_d$ \cite{bica_icassp16}. (b) Co-design system where only Rx are shared. In this \textit{joint multiple access} channel, the radar operates in monostatic mode and both systems transmit different waveforms that are orthogonal in spectrum, code or time \cite{mishra2017auto}. 
 		(c) In TX-shared co-design, the monostatic radar functions as a communications transmitter emitting a common JRC waveform \cite{muns2017beam}.
 		(d) A bi-static broadcast co-design with common TX, RX, and a joint waveform \cite{dokhanchi2019mmwave}. 
 		 The joint waveform transmitted by the TX vehicle bounces off from targets such as T1 and T2 and received by the Rx vehicle. A variant is in-band full duplex system with different waveforms but common TX and Rx \cite{donnet2006combining}. The term `BS' stands for `base station'.
}
 		\label{fig:jrc_top}
 \end{figure}
The existing mmWave JRC systems could be classified by the joint use of the channel \cite{bliss_RF_convergece_17,geng2018fusion} (Fig.~\ref{fig:jrc_top}). In the \textit{spectral coexistence} approach, radar and communications operate as separate entities and focus on devising strategies to adjust transmit parameters and mitigate the interference adaptively for the other \cite{cohen2017spectrum}.
To this end, some information exchange between the two systems, i.e. spectral cooperation, may be allowed but with minimal changes in the standardization, system hardware and processing. In \textit{spectral co-design} \cite{dokhanchi2019mmwave,bliss_RF_convergece_17},
new \textit{joint} radio-frequency sensing and communications techniques are developed where a single unit is employed for both purposes while also accessing the spectrum in an opportunistic manner. New fully-adaptive, software-defined systems are attempting to integrate these systems into same platform to minimize circuitry and maximize flexibility.
Here, each transmitter and receiver may have multiple antennas in a phased array or Multiple-Input Multiple-Output (MIMO) 
configuration. In the next section, we discuss mmWave systems  based on co-existence and follow it by co-design methods in Section~\ref{sec:jrc}.

\section{JRC at mmWave: Coexistence}
\label{sec:individual}
Interference management is central to spectral coexistence of different radio systems. This, typically requires sensing the state of the shared spectrum and adjusting TX and RX parameters so that the impact of interference is sufficiently reduced and individual system performance is enhanced. We now present the figures of merit qualifying system performance and then discuss methodologies for mmWave coexistence.

\subsection{Communications Performance Criteria}
\label{subsec:comm_crit}
Since the goal of communications systems is to transfer data at a high rate error-free for a given bandwidth, the commonly used performance criteria include quality of service (QoS) indicators such as spectral efficiency, mutual information, channel capacity, pairwise error probability, bit/symbol error rates (BER/SER), and 
signal-to-interference-and-noise ratio (SINR). Given a communications signal model, the achievable spectral efficiency can be used as a universal communications performance criterion. In practice, the achievable spectral efficiency $r$ is an upper bound, while the effective spectral efficiency $r_\eff$ depends on the implemented receiver (e.g. minimum mean square error or MMSE \cite{ShiZhoYao:Capacity-of-single-carrier:04}, decision feedback \cite{TakKyrHan:Performance-evaluation-of-60-GHz-radio:12} or time-domain equalizer \cite{liu2013digital}), and is a fraction of the achievable spectral efficiency. The effective communications rate is then the product of the signal bandwidth $W$ and $r_\eff$.
%

\subsection{Radar Performance Criteria}
\label{subsec:rad_crit}
Radar systems, by virtue of their use in both detection and estimation, lend themselves to a plethora of performance criteria depending on the specific task. Target detection performance is characterized by probabilities of correct detection, mis-detection, and false alarm. 
In parameter estimation task, mean square error (MSE) or variance in comparison to the Cram\'er-Rao Lower Bound (CRLB) is commonly considered. The CRLB defines the lower bound for estimation error variance for unbiased estimators. 
There are also several radar design parameters such as range/Doppler/angular resolution/coverage and the number of targets a radar can simultaneously resolve. In particular, the radar's ability to discriminate in both range and velocity is completely characterized by the \textit{ambiguity function} (AF) of its transmit waveform; it is obtained by correlating the waveform with its Doppler-shifted and delayed replicas.

\subsection{Interference Mitigation}
The mmWave radar and communications TX and RX can use all of their degrees of freedom (DoFs) such as different antennas, frequency, coding, transmission slots, power, or polarization to mitigate or avoid mutual interference. 
Interference may also be caused by leakage of signals from adjacent channels 
because of reusing identical frequencies in different locations. 
In general, higher the frequency in mmWave bands, weaker the multipath effects. The transmitters can adjust their parameters so that the level of interference is reduced at the receiver. To this end, awareness about the dynamic state of the radio spectrum and interference experienced in different locations, subbands and time instances is desired. This may be in the form of feedback provided by the receivers to the transmitter about the channel response and SINR. Both the TX and RX can be optimized such that the SINR is maximized at the receivers for both subsystems.  

\subsubsection{Receiver Techniques}
Interference mitigation may be performed only at the RX rendering channel state information (CSI) exchange optional. Typically, this requires multiple antenna at RX, a common feature at mmWave, and processing of the received signals in spatial and/or temporal domain. These techniques employ receive array covariance matrix $\bm{\Sigma}$ (or its estimate $\hat{\bm{\Sigma}}$) in certain interference canceling RX structures. 
Here, the received signal space spanned by eigenvectors of $\bm{\Sigma}$ is divided into two orthogonal subspaces of signal and interference-plus-noise. The received signal is then projected to a subspace orthogonal to the interference-and-noise subspace to enable processing of practically interference-free signals. If the interference impinges the receiver from angles different than the desired signal, RX beamforming is commonly used \cite{geng2018fusion}. The beampattern design ensures high gains towards the desired signals and steers nulls towards the interference. Common solutions include Minimum Variance Distortion-less Response (MVDR), 
 Linearly Constrained Minimum Variance (LCMV) and diagonal loading \cite{vorobyov2014adaptive}. 

Advanced interference cancellation receivers 
estimate CSI, use feedback about channel response or sense other properties of the state of the radio spectrum. These estimates are later used to cancel the interference contribution from the overall received signal. The coherence time of the channels should be sufficiently long that the feedback or channel estimates are not outdated during the interference cancellation process. These techniques either require knowledge of modulation schemes employed by coexisting radio systems, or are applied to digital modulation methods only. A prime example is the Successive Interference Cancellation (SIC) 
method that decodes and subtracts the strongest signal first from the overall received signals and the repeats the same procedure by extracting the next weaker signal from the residual signal and so on \cite{bliss_RF_convergece_17}. 
In the absence of CSI, non-traditional radar interference models are used for robust communications signal decoders \cite{ayyar2019robust}.

\subsubsection{Transmitter Techniques}
Adapting transmitters and optimizing transmit waveforms may be used to minimize the impact of interferences in coexistence systems. 
In a radar-communications coexistence scenario, for example, the optimization objective could be maximizing the SINR at each receiver while providing desired data rate for each communications user and target Neyman-Pearson detector performance for radar users. Designing a precoder for each transmitter or/and decoders for each receiver achieves this goal by steering the interferences to different space than the desired signals. 

One such example design in the context of MIMO communications and MIMO radar is the Switched Small Singular Value Space Projection (SSSVSP) method \cite{Mahal15} in which the interference is steered to space spanned by singular vectors corresponding to zero or negligible singular values. This method requires information exchange between the radar subsystem and communications base-stations. Another example of a precoder-decoder design for interference management in radar-communications coexistence is via Interference Alignment (IA) \cite{cui_spawc18} where IA coordinates co-existing multiple transmitters such that their mutual interference aligns at the receivers and occupies only a portion of the signal space. The interference-free signal space is then used for radar and communications purposes. 

\section{JRC at mmWave: Co-Design}
\label{sec:jrc}
Central towards facilitating the co-design of radar and communications systems are waveform design and their optimization exploiting available DoFs (spatial, temporal, spectral, polarization). The optimization is based on the system performance criteria and availability of channel state information (CSI), awareness about target scene and the levels of unintentional or intentional interference at the receivers. 

\subsection{JRC Performance Criteria}
In co-design, JRC waveforms are modeled to simultaneously improve the  functionalities of both subsystems  with some quantifiable trade-off. 
In \cite{Bli:Cooperative-radar-and-communications:14}, a radar round-trip delay estimation rate is developed and coupled with the communications information rate. This radar estimation, however, is not drawn from the same class of distributions as that of communications data symbols and, therefore, provides only an approximate representation of the radar performance. However, potential invalidity of some assumuptions limits the extension of this to estimation of other target paramters. 

The mmWave designs in \cite{KumVorHea:VIRTUAL-PULSE-DESIGN:19,KumNguHea:Performance-trade-off-in-an-adaptive:17} for single- and multiple-target scenarios suggest an interesting JRC performance criterion which attempts to parallel the radar CRLB performance with a new effective communications symbol MMSE criteria 
as a function of effective maximum achievable communications spectral efficiency, $r_\eff$. The MMSE communications criteria here is analogous to the mean-squared error distortion in the rate distortion theory. 
Let $\mmse_\comm$ be the MMSE of a communications system with spectral efficiency $r$. Then $\mmse_\comm$ and $r$ are related to each other through the equation 
$\frac{1}{N}
\Tr {\log_2 {\mmse_\mcom}} = -r
$, where $N$ is the code length. Therefore, the effective communications distortion MMSE (DMSE) that satisfies 
$\frac{1}{N}
\Tr {\log_2 {\dmse_\eff}} =  {-r_\eff} = {-\delta \cdot r}$ can be defined as
$\dmse_\eff \triangleq  \mmse_c^\delta$,
where $\delta$ is a constant fraction of communications symbols transmitted in a CPI with the channel capacity $C$. The performance trade-off between communications and radar is quantified in terms of a weighted combination of the scalar quantities $\frac{1}{N}
\Tr{\log_2 \dmse_\eff}$ and $\frac{1}{Q}
\Tr{\log_2 \textrm{CRLB}}$, respectively, where the log-scale is used to achieve proportional fairness between the communications distortion and radar CRLB values and $Q$ is the number of detected targets. 
Pareto-optimal solutions that assign weights to different design goals have also been explored in this context \cite{ciuonzo2015pareto}.

Mutual information (MI) is also a popular waveform optimization criteria. At the radar receiver, depending on whether the communications signal reflected off the target is treated as useful energy or interference or ignored altogether, a different MI-based criterion results \cite{bica_icassp16}. 
Although MI maximization enhances the characterizing capacity of a radar system, it does not maximize the probability of detection. 
The optimal radar signals for target characterization and detection tasks are generally different \cite{bica_icassp16,cohen2017spectrum}.

\subsection{Radar-Centric Waveform design}
We first consider the appropriate radar-centric waveforms here. These range from conventional signals to emerging multi-carrier waveforms.
\paragraph{Conventional Continuous Wave and Modulated Waveforms}
A simple continuous-wave (CW) radar provides information about only Doppler velocity. To extract range information, either the frequency/phase of CW signal is modulated or very short duration pulses are transmitted. In practice, the well-known Frequency Modulated Continuous Wave (FMCW) and Phase Modulated Continuous Wave (PMCW) radars are used. A typical FMCW radar transmits one or multiple chirp signals wherein the frequency increases or decreases linearly in time and then the chirps reflected off the targets are captured at the receiver. 
Chirp bandwidth of a few GHz may be used to provide a range resolution of a few centimeters,  e.g, $4$ GHz chirp achieves a range resolution of $3.75$ cm. For PMCW, binary pseudorandom sequences with desirable autocorrelation/  cross-correlation properties are typically used.  The AF of PMCW has lower sidelobes than FMCW and PMCW is also easier to implement in hardware \cite{dokhanchi2019mmwave}. 

\begin{figure}[t]	
		\centering
\includegraphics[width=1.0\textwidth]{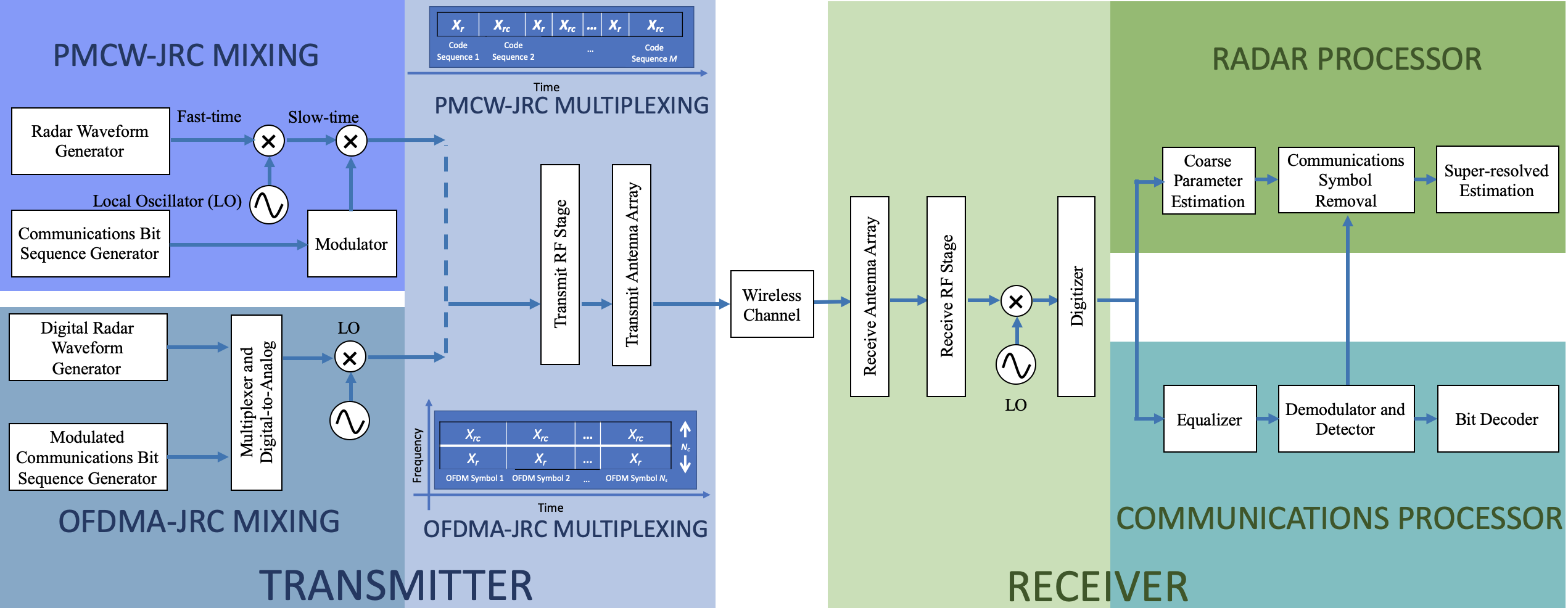}
		\caption{A simplified block diagram showing major steps of transmit and receive processing for a general mmWave JRC system. In case of PMCW-JRC, the radar and communications waveforms are combined in the analog hardware before the RF stage. On the other hand, the information bits from these two subsystems are mixed digitally in OFDMA-JRC. The multiplexing of radar-only and radar-communications frame for both PMCW- and OFDMA-JRC are depicted in the transmit portion. The receive processing for both systems is largely similar. }
		\label{fig:jrc_block}
\end{figure}

A general bi-static, uniform linear array (ULA) PMCW-JRC system \cite{dokhanchi2019mmwave} follows the topology shown in Fig.~\ref{fig:jrc_top}d. The transmitter sends $M$ repetitions of the PMCW code of length $L$ from each of its $N_{t}$ transmit antennas. The Doppler shift and flight time for the paths are assumed to be fixed over the CPI. The reflections from $Q$ targets impinge on $N_{r}$ receive antennas. Let $t_{c}$ be chip time (time for transmitting one element of one PMCW code sequence, i.e., fast-time). The Doppler shifts and the flight time for every path are assumed to be fixed over a coherent transmission time $Mt_b$, where $t_{b}=Lt_{c}$ is the time taken to transmit one block of code, i.e., slow-time. The transmit waveform takes the form,
	\begin{align} \label{2}
	x_i(t)=\sum_{m=0}^{M-1} \sum_{l=0}^{L-1} a_{m}e^{j\zeta_l} s(t-lt_{c}-mt_{b})e^{j2\pi f_{c}t} e^{j(i-1)k d \sin\beta}, 
	\end{align}
where $i\in[1,N_{t}]$ and $ a_{m}=  e^{j\phi_{m}} $ denote differential PSK symbols (DPSK) over slow time (time for sending one code sequence). The DPSK modulation is robust to constant phase shifts. Further, $s(t)$ is the elementary baseband pulse shape, $\zeta_{l} \in \{0, \pi \} $ is the binary phase code, $ e^{j(n-1)k d \sin\beta}$ is beam-steering weight for $n$th antenna, $k=\frac{2\pi}{\lambda}$ is wave number, and $\beta$ is angle between the radiating beam and the perpendicular to the ULA (for simplicity, we consider only azimuth and ignore common elevation angles). The transmitter steers the beam in multiple transmission from $[\dfrac{-\pi}{2}$, $\dfrac{\pi}{2}]$, each time with angle $\beta$. As shown in Fig.~\ref{fig:jrc_block}, the communications and radar waveform for PMCW-JRC are combined in analog hardware.

Let $\Delta V^{(1)}_{q}$ be the radial relative velocity between the transmitter and $q$th path, where superscript $(\cdot)^{(1)}$ refers to transmitter-target path, and the corresponding Doppler shift is $f^{(1)}_{D_{q}}= \frac{\Delta V^{{(1)}}_{q}}{c}f_c$, where $c=3\times 10^8$ m/s is the speed of light. The signal impinging on $q$th scatterer is,
\begin{align} \label{eq:signal@target}
\hspace*{-0.1in} 
z_{q,n}(t)= \sum_{m=0}^{M-1} \sum_{l=0}^{L-1} h^{(1)}_{q,n}a_{m}e^{j\zeta_{l}} s\big(  t-lt_{c}-mt_{b}-\tau^{(1)}_{q} \big)
 e^{j2\pi f_{c}t- j2\pi f^{(1)}_{D_{q}}t  - j 2\pi f_{c} \tau^{(1)}_{q}},
\end{align} 
where $\tau^{(1)}_{q}$ and $h^{(1)}_{q,n}$ are $q$th point scatterer time delay and propagation loss for each path, respectively. We exploit the standard narrowband assumption to express the received signal as a phase-Doppler shifted version of the transmit signal. Assume $\tau_{q}=\tau^{(1)}_{q}+\tau^{(2)}_{q}$ be the total flight time corresponding to a bi-static range $R_{q}=c\tau_{q}$, where superscript $(\cdot)^{(2)}$ denotes variable dependency on the target-receiver path. Assume $f_{D_{q}}=f^{(1)}_{D_{q}}+f^{(2)}_{D_{q}}$ to be the bi-static Doppler shift, and $\psi_{q}$ be the angle between the $q$th scatterer and perpendicular line to receive ULA. After TX/RX beamforming and frequency synchronization, the received signal at antenna $p$, obtained as a superposition of these reflections takes the form,
\begin{flalign} \label{4}
& 
\tilde y_{p}(t)= \sum_{q=1}^{Q}\sum_{n=1}^{N_{t}} h^{(2)}_{q,p}z_{q,,n}(t-\tau^{(2)}_{q})  
	e^{j2\pi f^{(2)}_{D_{q}}t} + \tilde N_p(t) \nonumber\\
&
=\sum_{q=1}^{Q}\sum_{n=1}^{N_{t}}\sum_{m=0}^{M-1} \sum_{l=0}^{L-1}
	h^{(2)}_{q,p} h^{(1)}_{q,n} a_{m} e^{j\zeta_{l}}  s(t-lt_{c}-mt_{b}-\tau^{(1)}_{q}-\tau^{(2)}_{q})
 e^{j2\pi (f_{c}-f^{(1)}_{D_{q}}-f^{(2)}_{D_{q}})t}  e^{j\eta_{q}} e^{-jk d \sin(\psi_{q}) (p-1)}+ \tilde N_p(t),
\end{flalign}
where $ e^{j\eta_{q}}= e^{-j2\pi \big( f_{c}(\tau^{(1)}_{q}+\tau^{(2)}_{q})+f^{(1)}_{D_{q}}\tau^{(2)}_{q} \big)}$ is a static phase shift, $h^{(2)}_{q,p}$ accumulates the effect of $q$th transmitter-target-receiver point scatterer, path-loss and RCS of the target, and $\tilde N_p(t)$ is complex circularly symmetric white Gaussian noise with variance $\sigma^2$. An extended target is modeled as a cluster of points. This combined with the superposition of reflections from independent scatterer renders the model in \eqref{4} applicable for extended targets. After downconversion to baseband and ignoring RCS dependency on Tx and Rx antennas, i.e., $\sum\limits_{n=1}^{N_t}h^{(1)}_{q,n}h^{(2)}_{q,p} e^{j\eta_{q}} = \sum_{n=1}^{N_t}\acute d_{q,p,n}=N_t \acute d_q = d_q$, received signal is
\begin{align}\label{PMCW-JRC_rx}
	y_{p}(t)= \sum_{q=1}^{Q}\sum_{ m=0}^{M-1} \sum_{l=0}^{L-1} d_{q} a_{m} e^{-j2\pi f_{D_q}t}c^{p-1}_q e^{j\zeta_{l}}s(t- lt_{c}- mt_{b}-\tau_{q}) + N_p(t) , \ p\in[1,N_{r}],
\end{align}
where $c_{q}= e^{-jk d \sin(\psi_{q})}$. Collecting the Nyquist time samples for the antenna $p$ and rearranging them accordingly to slow/fast-time, we form a matrix,
\begin{align} \label{eq:sys_mod1}
	\boldsymbol Y_p^{\text{PMCW-JRC}} = \sum_{q=1}^{Q} c_q^{p-1} d_q \text{Diag}\{\boldsymbol a\} \big[ ( \boldsymbol b_q^T \odot \boldsymbol s^T \boldsymbol P_{k_q}) \otimes \boldsymbol e_q \big] +\boldsymbol N_p \  \in \mathbb{C}^{M\times L},       
\end{align}
where vectors $ \boldsymbol{e}_q = [e^{j2\pi f_{D_q} m L t_{c}}]_{m=1}^{M}$ and $\boldsymbol{b}_q= [e^{j2\pi f_{D_q} l t_{c}}]_{l=1}^{L}$ collect Doppler samples in slow and fast time, respectively, $\boldsymbol{s}= [e^{j\zeta_l}]_{l=0}^{L}$ contains $L$ chips of code sequence, and $\boldsymbol P_{k_q}$ is a cyclic permutation matrix for a shift of ${k_q}$ as
\begin{align}
\boldsymbol P_{k_q} = \begin{bmatrix} 
\mathbf0_{K_q\times L-K_q} & \mathbf I_{K_q\times K_q}\\
\mathbf{I}_{L-K_q\times L-K_q} & \mathbf{0}_{L-K_q\times K_q}
\end{bmatrix} \in \mathbb{C}^{L \times L},
\end{align} 
where $k_q \in \{0,\cdots , L-1\}$ is determined by range of the $q$th scatterer. If there is no delay between transmitter and receiver for all paths, then $k_q=0$ for all $q$ and $\boldsymbol P_{k_q}$ becomes identity matrix.

In a PMCW-JRC, the communications symbols and Doppler parameters are coupled thus leading to a non-identifiable model. This is resolved by a multiplexing strategy through which unknown parameters in the received signal are uniquely identified. The PMCW-JRC adopts time-division multiplexing between radar-only ($\bm X_r$) and joint radar-communications ($\bm X_{rc}$) frames which are transmitted for $\mu$ and ($1-\mu$) \% of the CPI, respectively. The value of $\mu$ depends on the amount of prior knowledge about the target scene. As a case in point, when the scene is stationary such as driving a straight path on a highway, we may not need full sensing capacity and can scale up the allocated time appropriately for communications. A coarse estimate of radar target parameters (range, angle and Doppler) is obtained from $\boldsymbol Y_p^{\text{PMCW-JRC}}$ of radar-only frames $\bm X_r$ while communications symbols are extracted from the received signal samples of $\bm X_{rc}$ frame. After extracting communications symbols from $\bm X_{rc}$, the residual signal is exploited for further improving the radar target estimates through low-complexity JRC super-resolution algorithms \cite{dokhanchi2019mmwave}.

\paragraph{Multi-Carrier Waveforms}
Multi-carrier waveform radars provide additional DoFs to deal with dense spectral use and demanding mmWave target scenarios like drones, low-observable objects, and large number of moving vehicles in automotive scenario. Different DoFs can be used in an agile manner to achieve optimal performance depending on the radar task, nature of targets, and state of the radio spectrum. A general drawback of multi-carrier radar waveforms is their time-varying envelope leading to an increased Peak-to-Average-Power-Ratio (PAPR) or 
Peak-to-Mean-Envelope-Power-Ratio (PMEPR) which makes it difficult to use the amplifiers efficiently when high transmit powers are needed. However, in mmWave radars, the transmit powers tend to be small and surveillance ranges are short. The PAPR reduction is achieved by not allocating all subcarriers or by using appropriate coding/waveform design. Hence, the PAPR issue in mmWave may be less severe. 

Multi-carrier Complementary Phase Coded (MCPC) waveform \cite{levanon_mf_radar_2000}, wherein each subcarrier is modulated by a pseudorandom code sequence of a specific length, is also a viable mmWave JRC candidate. The MCPC design exploits DoFs in spectral and code domain. 
In a sense, it is related to OFDM because after each subcarrier is modulated by a code in time-domain, the subcarriers remain orthogonal without intercarrier interference. If the subcarriers are uncoded, the waveform is exactly OFDM. The inter-carrier spacing in MCPC needs to accommodate the spreading of the signals in frequency due to phase codes such as Barker, P3 or P4 polyphase codes \cite{skolnik2008radar}. This is achieved by choosing the inter-carrier spacing to be inverse of the chip duration. In OFDM, intercarrier spacing is smaller. A Generalized Multi-carrier Radar (GMR) waveform devised in \cite{bica_ciss_14,bica_trsp_16} subsumes most of the widely used radar waveforms such as pseudo random frequency hopping (FH), MCPC, OFDM and linear step approximations of linear FM signals, as special cases. A matrix model of transmitter and receiver is developed for GMR that allows for defining the waveforms and codes, spreading in time and frequency domain, power allocations and active subcarriers using a compact notation. Different waveforms are obtained by choosing the dimensions of the matrix model and filling the entries appropriately. This approach allows for relaxing perfect orthogonality requirement; 
this may lead to a better resolution of target delays and Doppler velocities at mmWave. 

\paragraph{Spatial DoFs and Multiple Waveforms}
A few different solutions use the same waveform for both subsystems but make use of radar's spatial DoFs for communications symbols. For instance, in \cite{Hassan16}, the radar array beampattern sidelobes are modulated by communications messages along user directions. In \cite{Hassan18}, the communications symbols are represented by different pairing of antennas and waveforms in a MIMO configuration. Spatial DoFs are also useful for adaptively canceling specific users. A joint beamforming method is suggested in \cite{hassanien_spawc18} for a dual-function radar-communications (DFRC) that comprises MIMO radar and communications systems assuming full-duplex transmission. The downlink communications signal is embedded into the transmit radar waveform and uplink communications takes place when the radar is in listening mode. This necessitates accurate synchronization among the subsystems. The technique utilizes spatial diversity by enforcing the spatial signature of the uplink signals to be orthogonal to the spatial steering vectors associated with the radar target returns. The receiver beamformer employs adaptive and non-adaptive strategies to separate the desired communications signal from echoes of targets, clutter, and noise even if they impinge the array from the same direction. Other solution paths consist of finding spatial filters to mitigate in-band MIMO communications interference through optimization of the sidelobe and cross-correlation levels in MIMO radar systems \cite{Aittom17,LiPetr16}, exploiting co-array processing with multiple waveforms \cite{ZhangVor15} and designing precoders/decoders through interference alignment \cite{CuiKoivu17}.

However, for mmWave JRC systems, the full-resolution ADCs at the baseband signal result in an unacceptably high power consumption. This makes it infeasible to utilize an RF chain for each antenna element implying that the prevailing MIMO systems that employ fully digital beamforming are not practical for mmWave systems. Thus, the benefits of using multiple waveforms for spatial mitigation in mmWave JRC systems are yet to be carefully evaluated. Currently, a single data stream model that supports analog beamforming with frequency flat TX/RX beam steering vectors is more common \cite{KumChoGon:IEEE-802.11ad-based-Radar::17}. Use of large antenna arrays in mmWave suggests that a feasible JRC approach could be to simply partition the arrays for radar and communications functionalities \cite{mishra2018sub}.

\subsection{Communications-Centric Waveform design}
The most popular communications signal for mmWave JRC is OFDM 
because it provides a stable performance in multipath fading and relatively simple synchronization \cite{donnet2006combining}. Also, frequency division in duplexing has an added advantage; unlike time-division duplexing, the former employs different bands for uplink and downlink so that the impact on the interference in radar systems is less severe. Some solutions \cite{donnet2006combining,dokhanchi2019mmwave} also employ the related Orthogonal Frequency Division Multiple Access (OFDMA) waveform for a JRC system. While the OFDM users are allocated on only time domain, the OFDMA users can be differentiated by both time and frequency. The latter, therefore, provides DoFs in both temporal and spectral domains. Although OFDM-JRC offers high dynamic range and efficient receiver processing implementation based on fast Fourier transform (FFT), it requires additional processing to suppress high side-lobes in receiver processing and reduce PAPR. Further, the OFDM cyclic prefix (CP) used to
transform frequency selective channel to multiple frequency flat channels leading to a simplified equalizer, may be a nuisance in the radar context. The CP may adversely affect the radar's ability to resolve ambiguities in radar ranging. Its length depends on number of channels, particularly the maximum excess delay that the radar signal may experience (time difference between first and last received component of the signal). For radar applications, the CP duration should be equal to or longer than the total maximum signal travel time between the radar platform and target. Other communications waveforms proposed for mmWave automotive JRC include spread spectrum, noise-OFDM, and multiple encoded waveforms \cite{dokhanchi2019mmwave}. We now examine mmWave OFDMA-JRC in detail.
\paragraph{OFDMA-JRC} Consider the same bi-static scenario of Fig.~\ref{fig:jrc_top}d that we earlier analyzed for the PMCW-JRC system. The OFDMA-JRC transmitter (Fig.~\ref{fig:jrc_block}) sends $N_s$ OFDM symbols from $N_{t}$ transmit antennas and reflections from $Q$ targets impinge on $N_{r}$ receive antennas. Assume that $\beta$ is angle of departure. The Doppler shift and flight time for the paths are assumed to be fixed over a CPI, i.e., $N_sT_{sym}$, where $T_{sym}$ is the duration of one OFDM symbol and $a_{n,m}$ are multiplexed communications/radar DPSK on $n$th carrier of $m$th OFDM symbol. Let $N_c$ be the number of subcarriers and $\Delta f$ be the subcarriers spacing, then the joint transmit waveform in baseband neglecting the CP is,
\begin{align} \label{MODEL_TX}
	& x_i(t)= \sum_{m=0}^{N_s-1} \sum_{n=0}^{N_c-1} a_{n,m} e^{j2\pi f_nt}
	e^{j k \sin(\beta) (i-1) \frac{\lambda}{2}} s(t-mT_{\text{sym}}),
\end{align}
where $s(t)$ is a rectangular pulse of the width $T_{\text{sym}}$, $i\in[1,N_t]$, $n$ and $m$ are frequency and time indices respectively, and $f_n=n\Delta f=\frac{n}{T_{sym}}$ \cite{dokhanchi2019mmwave}.	The received signal at the $p$th receiver over a CPI is,
\begin{align} \label{MODEL}
\tilde y_p(t)= \sum_{m=0}^{N_s-1} \sum_{q=1}^{Q} \sum_{n=0}^{N_c-1} \sum_{i=1}^{N_t}  d_{q,i,p} a_{n,m}e^{j2\pi f_n(t-\tau_{q})} e^{j2\pi f_{D_{q}} t} e^{j k \sin(\psi_{q}) (p-1) \frac{\lambda}{2}} s(t-m T_{sym}-\tau_{q}) + \tilde{N}_p(t), 
\end{align}
where $\tilde{N}_p(t)$ is the additive noise on antenna $p$, Similar to PMCW-JRC, $d_{q,i,p}$ denotes path-loss, phase-shift caused by carrier frequency  and RCS  of  the  target; $d_{q,i,p}$ is independent of the subcarrier index due to narrowband assumption. Similarly, the Doppler is assumed to be identical for all subcarriers given a small inter-carrier spacing. For notational convenience, we omit the noise in the following. We sample (\ref{MODEL}) at intervals $t_s=\dfrac{1}{N_c \Delta f}$ as,
\begin{align}  \label{ofdm_final0}
& 
\tilde y_p[t_s]
= \sum_{m=0}^{N_s-1} \sum_{q=1}^{Q} \sum_{n=0}^{N_c-1}  d_{q}  s_{n,m}
	e^{j2\pi \frac{nl}{N_c}}s(lt_s-m T_{\text{sym}}- \tau_{q}),
\end{align}
where $l\in[1,L]$, $n\in[1,N_c]$ and $L \leq N_c$, $d_{q}=\sum_{i=1}^{N_t}  d_{q,i,p}$ as before, and $\tilde s_{n,m}= a_{n,m} e^{-j2\pi n\Delta f \frac{R_{q}}{c}} e^{j2\pi m T_{\text{sym}}f_{D_{q}}} e^{j \pi \sin(\psi_{q}) (p-1)}$ $\tilde s_{n,m}$ contains information about range, Doppler, angle of arrival and communications. We assume the number of inverse Fast Fourier Transform (IFFT) points $N_c$ is equal to the number of fast-time samples $L$ in each OFDM symbol. The received signal samples can be viewed as a radar data cube in spatial, spectral and temporal domains with $N_t$ antennas, $N_c$ subcarriers and $N_s$ OFDM symbols. Let us stack the entire DPSK symbols into a matrix $\bm A \in \mathbb{C}^{N_c\times N_s} $ and $\bm a_m = [\bm A]_m$ be the communications symbols over all subcarriers at $m$th OFDM symbol time. For a given OFDM symbol, say $m$, collecting signals from all subcarriers across different antennas leads to the following slow-time slice of the data cube
	\begin{align} \label{time_slice_ofdm}
	\bm Y_{m}^{\text{ OFDMA-JRC}}
	= \bm F_{N_c} Diag(\bm{a}_m ) \boldsymbol \Xi( \frac{-\Delta fR_q}{c}) Diag(\boldsymbol d) \boldsymbol C \ \in \mathbb{C}^{N_c\times N_r},
	\end{align}
where $m\in[1,N_s]$, $ \bm \Xi( \frac{-\Delta fR_q}{c}) =[e^{-j2\pi n\Delta f \frac{R_q}{c}}]_{n=1,q=1}^{N_c,Q} \in \mathbb{C}^{N_c\times Q}$, $\bm C =[e^{j k \sin(\psi_q) (p-1) \frac{\lambda}{2}} ]_{q=1,p=1}^{Q,N_r} \in \mathbb{C}^{Q\times N_{r}}$ and $\bm d=\begin{bmatrix} d_1 & \cdots & d_Q \end{bmatrix}$. Further, $\bm F_{N_c} = [e^{j2\pi \frac{nl}{N_c}}]_{l=0,n=0}^{N_c-1,N_c-1}$ denotes $N_c$-point IFFT matrix. To estimate Doppler shifts, we consider subcarrier slice of data cube \eqref{ofdm_final0}:
	\begin{align}  \label{eq:fd_est}
	\bm Z_{n}^{\text{OFDMA-JRC}}= Diag(\bm a_n) \bm \Xi(f_{D_q}T_{sym}) Diag(\bm d) \bm C \in \mathbb{C}^{N_s\times N_r},
	\end{align}
where $\bm a_n=[\bm A]_n\in \mathbb{C}^{N_s}$ are the DPSK symbols over slow-time, \bm $\bm \Xi(f_{D_q}T_{sym})=[e^{j2\pi m T_{sym}f_{D_q}}]_{m=1,q=1}^{N_s,Q}$.
	
As in PMCW-JRC, the receive processing of OFDMA-JRC is affected by coupling of communications symbols with a radar parameter (range in case of OFDMA-JRC). To ensure that range estimation does not suffer by using all subcarriers, frequency-division multiplexing is employed (\ref{fig:jrc_block}) such that $\mu$\% of the OFDMA subcarriers are allocated to radar (with known $a_{n,m}$ on these subcarriers) and the rest to JRC. The rest of the OFDMA-JRC receive processing is similar to PMCW-JRC (Fig.~\ref{fig:jrc_block}) \cite{dokhanchi2019mmwave}.

\begin{figure}[t]	
		\centering
\includegraphics[width=1.0\textwidth]{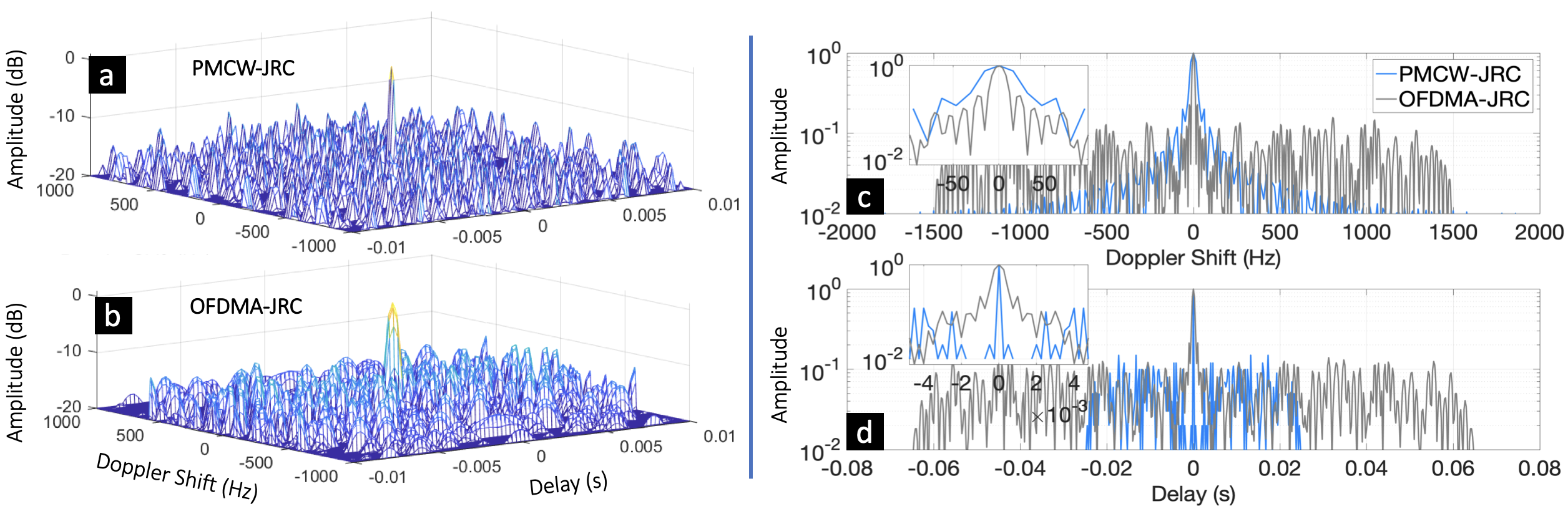}
		\caption{The AFs of bi-static mmWave JRC using (a) OFDMA (b) PMCW signals with the (c) Doppler and (d) delay cuts \cite{dokhanchi2019mmwave}. }
		\label{fig:amb_fcns}
\end{figure}

\begin{figure}[t]
		\centering
		\includegraphics[width=1.0\textwidth]{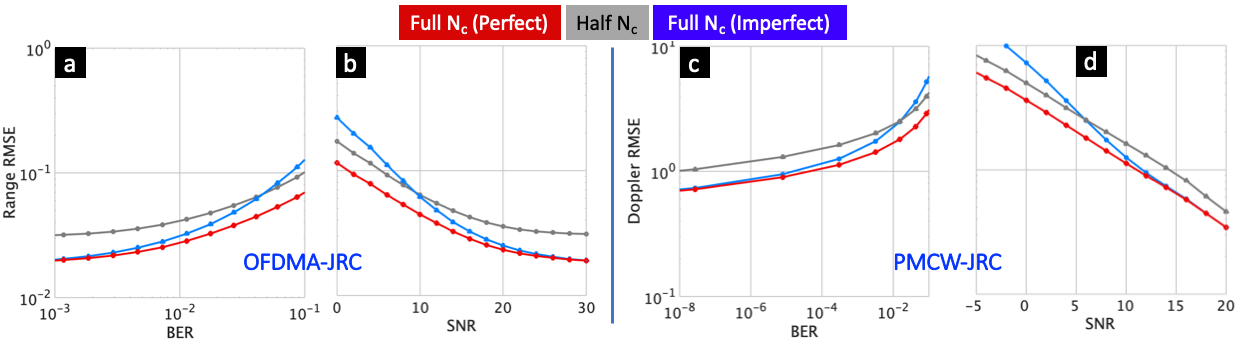}
		\caption{The root-mean-square-error (RMSE) of estimated range of a single target using OFDMA-JRC with respect to (a) SNR and (b) BER using half ($\mu$=50\%) or all subcarriers (full $N_c$) with perfect and imperfect recovery of communications symbols. The RMSE in Doppler estimate of a single target for PMCW-JRC using all and half frames with respect to (c) SNR and (d) BER. In both cases, JRC super-resolution algorithms \cite{dokhanchi2019mmwave} have been employed.}
		\label{fig:ofdm_pmcw_res}
\end{figure}

\paragraph{Comparison of PMCW- and OFDMA-JRC}
While OFDMA encodes radar and communications simultaneously in the entire \textit{time and space}, the PMCW does so in the entire \textit{frequency and space}; hence, their DoFs and design spaces are in different domains. While it turns out that the receive system models of both waveforms are mathematically identical after matched filtering and retrieve all JRC parameters using similar super-resolution algorithms \cite{dokhanchi2019mmwave,Dokhanchi2017Joint}, their individual performances mimic the respective communications and radar-centric properties. For example, the AF of the bi-static PMCW-JRC inherits the low sidelobes from its parent stand-alone PMCW radar waveform as shown in a comparison with the AF of OFDMA-JRC in Fig.~\ref{fig:amb_fcns}, given the same bandwidth. 
On the other hand, the PMCW-JRC is more sensitive to the number of users while the orthogonality of waveforms in OFDMA-JRC makes the latter robust to inter-channel interference. 
Finally, in a networked vehicle scenario, it requires less complex infrastructure and processing to apply PMCW with predefined or stored sequences rather than using OFDMA to adaptively allocate band to each user \cite{dokhanchi2019mmwave,donnet2006combining}. A comparison of estimation errors in the coupled parameter - range for OFDMA-JRC and Doppler for PMCW-JRC - using JRC super-resolution recovery \cite{dokhanchi2019mmwave} is shown in Fig.~\ref{fig:ofdm_pmcw_res} for $\mu=50$\%. 

\subsection{Joint Coding}
Recently, existing mmWave communications protocols that are embedded with codes which exhibit favorable radar ambiguity functions are garnering much attention for JRC. In particular, the 60 GHz IEEE 802.11ad wireless protocol has been employed with time-division multiplexing of radar-only and radar-communications frame. In general, these designs have temporal DoF (for a monostatic radar case). The IEEE 802.11ad single-carrier physical layer (SCPHY) frame consists of a short training field (STF), a channel estimation field (CEF), header, data and beamforming training field. 
The STF and CEF together form the SCPHY preamble.  CEF contains two 512-point sequences $Gu_{512}[n]$ and $Gv_{512}[n]$, each containing a \textit{Golay complementary pair} of length $256$, $\{Gau_{256}, Gbu_{256}\}$ and $\{Gav_{256}, Gbv_{256}\}$, respectively. 
A Golay pair has two sequences $Ga_N$ and $Gb_N$ each of the same length $N$ with entries $\pm1$, such that the sum of their \textit{aperiodic} autocorrelation functions has a peak of $2N$ and zero sidelobes: 
\begin{align}
Ga_N[n]*Ga_N[-n] + Gb_N[n]*Gb_N[-n] = 2N\delta[n],     
\end{align}
where $*$ denotes linear convolution. This property is useful for channel estimation and target detection. 

\begin{figure}[t]
\centering 
\includegraphics[width=1.00\textwidth]{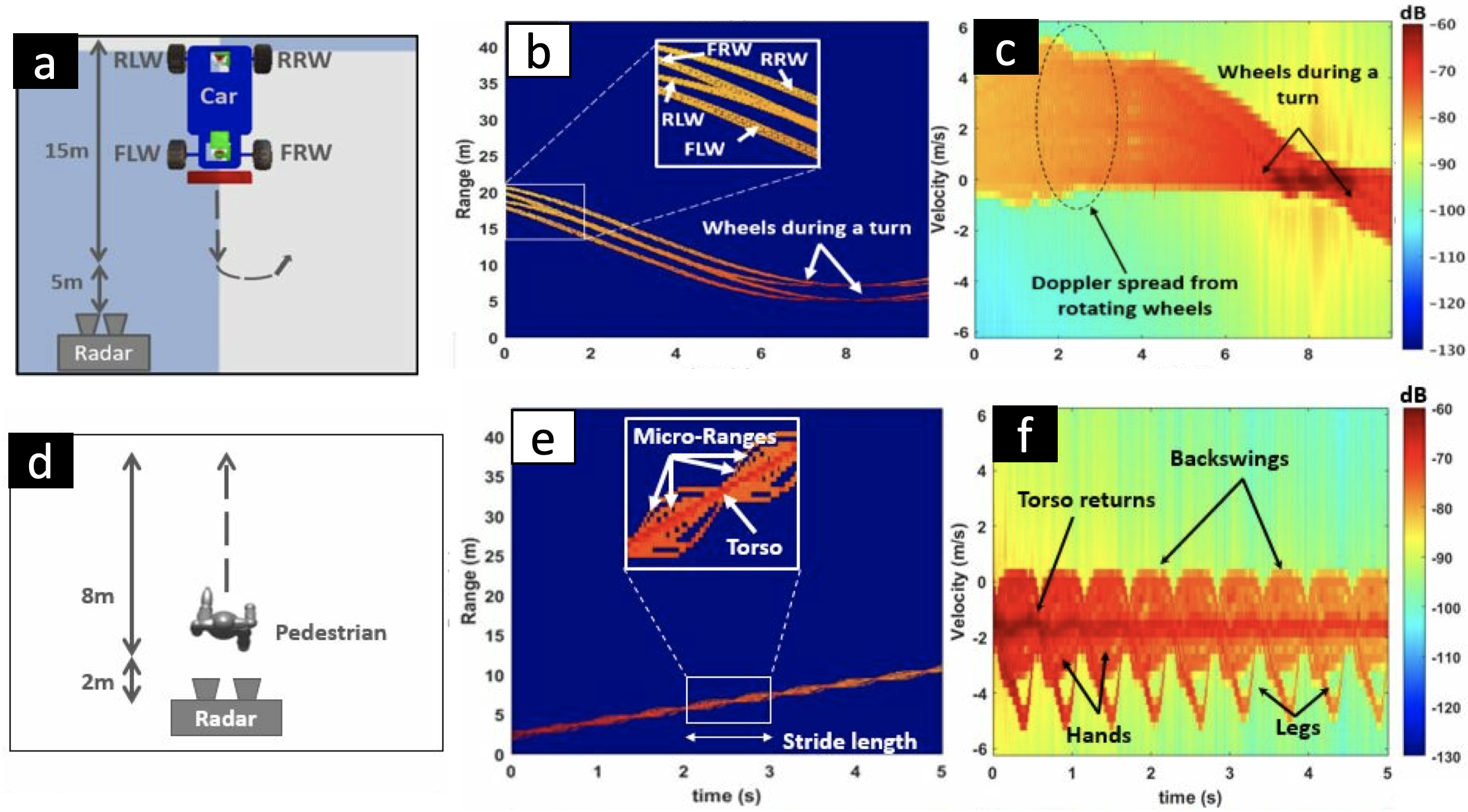}
\caption{Radar signatures generated from animation models of (a) a small car and (b) a pedestrian using Doppler-resilient 802.11ad waveform \cite{duggal2019micro, mishra2019doppler}. As the targets move radially in front of the radar on the marked trajectories, the movements of the front right, front left, rear right, and rear left wheels (FRW, FLW, RRW and LLW, respectively) of the car as well as the torso, arms, and legs of the pedestrian are individually observed in (b, e) range-time and (c, f) Doppler-time domains.}
\label{fig:dop_res_sign}
\end{figure}

By exploiting the preamble of a single SCPHY frame for radar, the existing mmWave 802.11ad waveform simultaneously achieves a cm-level range resolution and a Gbps data rate \cite{KumChoGon:IEEE-802.11ad-based-Radar::17}. The limited velocity estimation performance of this waveform can be improved by using multiple fixed length frames in which preambles are reserved for radar \cite{KumChoGon:IEEE-802.11ad-based-Radar::17}. While this increases the radar integration duration leading to more accurate velocity estimation, the total preamble duration is also prolonged causing a significant degradation in the communications data rate \cite{KumNguHea:Performance-trade-off-in-an-adaptive:17}. A joint coding scheme based on the use of sparsity-based techniques in the time domain can minimize this trade-off between communications and radar \cite{KumVorHea:VIRTUAL-PULSE-DESIGN:19}. Here, the frame lengths are varied such that their preambles (exploited as radar pulses) are placed in non-uniformly. These non-uniformly pulses in a CPI are then used to construct a virtual block of several pulses increasing the radar pulse integration time and enabling an enhanced velocity estimation performance. If the channel is sparse, the same can be achieved in frequency-domain using sub-Nyquist processing \cite{mishra2017sub}. In \cite{mishra2019doppler}, the wide bandwidth of mmWave is exploited using a Doppler-resilient 802.11ad link to obtain very high resolution profiles in range and Doppler with the ability to distinguish various automotive targets. Fig.~\ref{fig:dop_res_sign} shows distinct, detailed movements of each wheel of a car and body parts of a pedestrian as detected by an 802.11ad-based Doppler-resilient short range radar.

\subsection{Carrier Exploitation}
\begin{figure}[t]	
		\centering
\includegraphics[width=1.0\textwidth]{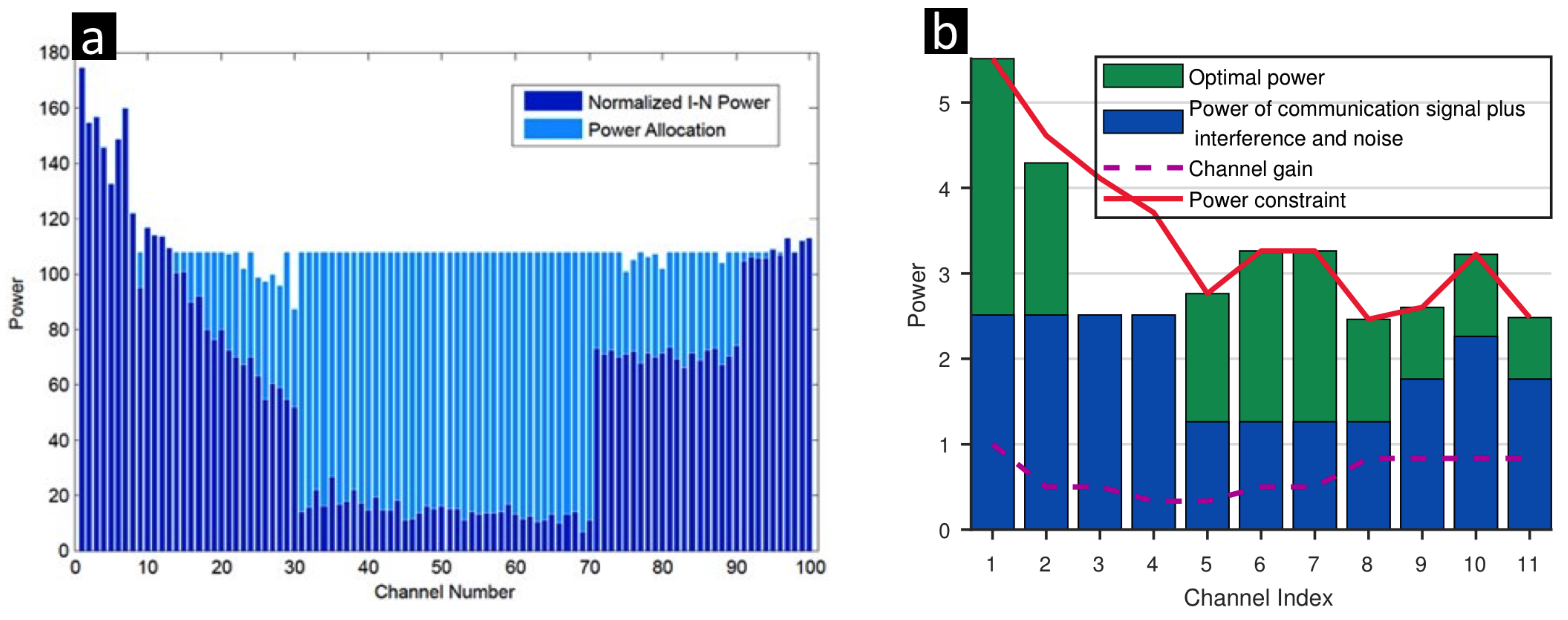}
		\caption{Power allocation solutions for JRC carrier exploitation via (a) water-filling and (b) Neyman-Pearson test \cite{bica2015opportunistic}.}
		\label{fig:wf}
\end{figure}
Selecting active subcarriers and controlling their power levels or PAPR in an adaptive manner is also useful for interference management. Radar systems generally utilize entire bandwidth to achieve high resolution. On the other hand, communications systems often allocate resource blocks of certain number of subcarriers to each user based on channel quality indicator (CQI) to satisfy their rate and system QoS requirements.  Through feedback from the receivers, spectrum sensing, databases or other sources, the transmitters of both systems can have information about occupancy of different subcarriers, instantaneous or desired SINR levels, channel gains, and power constraints imposed by other coexisting subsystems. This awareness can be exploited in adaptively optimizing the power allocation among different subcarriers. An example of optimizing subcarrier power ($P_k$) allocations and imposing minimum desired rate constraints on wireless communications users and maximum power constraint $P_\mathrm{T}$ for the radar is as follows:
\begin{align}
\label{eq:c3_opt}
&\underset{P_k,\eta}{\text{maximize}}\;\; p_{\mathrm{D}}\nonumber\\
&\text{subject to} \;p_{\mathrm{FA}} \leq \alpha,\nonumber\\
&\;\;\;\;\;\;\;\;\;\;\;\;\;\;\;\log \left( {1+ SINR_k}  \right) \geq t_k, \; \forall k,\nonumber\\
&\;\;\;\;\;\;\;\;\;\;\;\;\;\;\;{\displaystyle \sum_{k=0}^{N-1}} P_k \leq P_\mathrm{T},
\end{align}
where $\eta$ is detection threshold for likelihood ratio test using Neyman-Pearson detection strategy with false alarm constraint $\alpha$.
Two example power allocations from the radar perspective are depicted in Fig.~\ref{fig:wf}. A water-filling solution (Fig.~\ref{fig:wf}a) obtained by maximizing Mutual Information between received data and target and channel response allocates the radar power to those parts of spectrum where the signal experiences the least attenuation and interference level is low. The second approach (Fig.~\ref{fig:wf}b) takes into account channel gains, required SINR values at communications subsystems while maximizing the radar performance in the Neyman-Pearson sense for target detection task.

\section{Cognition and Learning in mmWave JRC}
Some more recent enabling architectures and technologies for mmWave JRC where the system can sense, learn and adapt to the changes in the channel are as follows.
\paragraph{Cognitive Systems}
Cognitive radars and radios sense the spectrum and exchange information to build and learn their radio environment state. This typically implies channel estimation and feedback on channel quality. Spectrum cartography methods, that generate a map of spectrum access in different locations and frequencies at different time instances, have been developed in this context \cite{giannakis_cartography_2011}. Based on the obtained awareness, operational parameters of transmitters and receivers in each subsystem are adjusted to optimize their performance \cite{cohen2017spectrum}. 
Channel coherence times should be long enough for JRC to apply cognitive actions. Since this duration is in nanoseconds for mmWave environments, compressed sensing-based solutions aid in reducing required samples for cognitive processing \cite{mishra2017sub,mishra2017performance}.
\paragraph{Fast Waveforms}
Algorithms that develop cognitive waveforms should have low computational complexity in order to re-design waveforms on-the-fly, typically within a single CPI. This is especially important for mmWave systems where the fast-time radar waveform can easily have a length of tens of thousands samples. 
In \cite{LiVor18a}, waveform design in spectrally dense environment does not exceed a quadratic complexity. In \cite{mishra2017auto,mishra2017sub}, the mmWave radar based on sub-Nyquist sampling adaptively transmits in disjoint subbands and the vacant slots are used by vehicular communications.
\paragraph{Machine Learning}
In order to facilitate fast configuration of mmWave JRC links with low latency and high efficiency, machine learning is useful to acquire situational awareness. This implies learning the evolution of spectrum state over time (including classifying radar target responses or other waveforms occupying the spectrum), acquiring the channel responses, identifying underutilized spectrum and exploiting it in an opportunistic manner. The deep learning methods are widely applied for tasks such as target classification, automatic waveform recognition and determining optimal antennas and RF chains \cite{elbir2019deep}. Optimal policies for coexisting systems may be learned using reinforcement learning approaches like partially observable Markov decision process (POMDP) and restless multiarm bandit (RMAB) \cite{lundenSPM}.
\paragraph{Game Theoretic Solutions}
The interaction between radar and 
communications systems sharing spectrum can be analyzed from a game theory perspective \cite{mishra2019power}. The two systems or players form an adversarial, non-cooperative game because of conflicting interests in sharing the spectrum. The game is also dynamic due to continuously evolving spectral states over time. The utility function is designed to reflect the possible strategies based on the respective players' requirements. The solutions result in Nash or Stackelberg equilibrium which are the game states with the property that none or one of the players can do better, respectively. In comparison to sub-6 GHz, the solution space for mmWave is several GHz wide with much lower maximum transmit power.

\section{Summary}
We outlined various aspects of implementing JRC systems at mmWave. The sheer number of mmWave antennas and huge bandwidth pose new challenges in waveform design and receiver processing that was not seen in other bands. The dynamic and highly variable environments of mmWave applications require continuous cognition of the mmWave channel by both radar and communications. While there are still many open problems in this area, 
mmWave JRC is a precursor to an emerging frontier of sub-mmWave or THF JRC where THF communications 
would coexist with the promising technology of low-THF ($.1$-$1$ THz) automotive and imaging radars. 

\section*{Acknowledgements}
This work is partially funded by the European Research
Council grant titled Actively Enhanced Cognition based
Framework for Design of Complex Systems and Luxembourg
National Research Fund project Adaptive mmWave
Radar Platform for enhanced Situational Awareness: Design
and Implementation.

\bibliographystyle{IEEEtran}
\bibliography{main}

\end{document}